\documentstyle[epsf,aps]{revtex}

\begin{document}

\title{\null\vspace*{-1truecm}  \hfill\mbox{\small IISc-CTS-2/00}\\
            \vspace*{-0.2truecm}\hfill\mbox{\tt\small quant-ph/0002037}\\
\vspace*{0.2truecm} Quantum Algorithms and the Genetic Code}

\author{Apoorva Patel}
\address{CTS and SERC, Indian Institute of Science, Bangalore-560012, India\\
         E-mail: adpatel@cts.iisc.ernet.in}
\date{{\small\sl Invited lectures presented at the Winter Institute on
``Foundations of Quantum Theory and Quantum Optics'',\\
1-13 January 2000, S.N. Bose National Centre for Basic Sciences, Calcutta,
India. To appear in the proceedings.}}

\maketitle


\begin{abstract}
Replication of DNA and synthesis of proteins are studied from the view-point
of quantum database search. Identification of a base-pairing with a quantum
query gives a natural (and first ever!) explanation of why living organisms
have 4 nucleotide bases and 20 amino acids. It is amazing that these numbers
arise as solutions to an optimisation problem. Components of the DNA structure
which implement Grover's algorithm are identified, and a physical scenario
is presented for the execution of the quantum algorithm. It is proposed that
enzymes play a crucial role in maintaining quantum coherence of the process.
Experimental tests that can verify this scenario are pointed out.
\end{abstract}

\section{\bf Genetic Information}

I am going to talk about processes that form the basis of life and evolution.
The hypothesis that living organisms have adapted to their environment,
and have exploited the available material resources and the physical laws
governing them to the best of their capability, is the legacy of Charles
Darwin---survival of the fittest. This is an optimisation problem, but it
is not easy to quantify it in mathematical terms. Often we can explain
various observed features of living organisms \cite{lifeshapes}.
The explanation becomes more and more believable, as more and more of its
ingredients are verified experimentally \cite{chirality}.
Yet even when definite predictions exist, an explanation is an explanation
and not a proof; there is no way we can ask evolution to repeat itself and
observe it like many common scientific experiments.

With this attitude, let us look at life. Living organisms try to perpetuate
themselves \cite{perpetuation}.
The disturbances from the environment, and the damage they cause, make it
impossible for a particular structure to survive forever. So the perpetuation
is carried out through the process of replication. One generation of organisms
produces the next generation, which is essentially a copy of itself. The
self-similarity is maintained by the hereditary information---the genetic
code---that is passed on from one generation to the next. The long chains of
DNA molecules residing in the nuclei of the cells form the repository of
the genetic information. These DNA molecules control life in two ways:
(1) their own highly faithful replication, which passes on the information
to the next generation (each life begins as a single cell, and each cell in
a complex living organism contains identical DNA molecules), and (2) the
synthesis of proteins which govern all the processes of a living organism
(haemoglobin, insulin, immunoglobulin etc. are well-known examples of
proteins) \cite{protsynth}.

Computation is nothing but processing of information. So we can study what
DNA does from the view-point of computer science. In the process of designing
and building modern computers, we have learnt the importance of various
software and hardware features. Let us look at some of them.

The first is the process of digitisation. Instead of handling a single variable
covering a large range, it is easier to handle several variables each spanning
a smaller range. Any desired accuracy can be maintained by putting together
as many as necessary of the smaller range variables, while the instruction
set required to manipulate each variable is substantially simplified.
This simplification means that only a limited number of processes have to be
physically implemented, leading to high speed computation. Discretisation also
makes it possible to correct small errors arising from local fluctuations.
There are disadvantages of digitisation in terms of increase in the depth
of calculation and power consumption, but the advantages are so great that
digital computers have pushed away analogue computers to obscurity. Even
before the discovery of DNA, Erwin Schr\"odinger had emphasised the fact that
an aperiodic chain of building blocks carries information \cite{schrodinger},
just like our systems of writing numbers and sentences. The structure of DNA
and protein reveals that life has indeed taken the route of digitising its
information. DNA and RNA chains use an alphabet of 4 nucleotide bases, while
proteins use an alphabet of 20 amino acids.

The second is the packing of the information. When there are repetitive
structures or correlations amongst different sections of a message, that
reduces its capacity to convey new information---part of the variables are
wasted in repeating what is already conveyed. Claude Shannon showed that
the information content of a fixed length message is maximised when all the
correlations are eliminated and each of the variables is made as random as
possible. Our languages are not that efficient; we can immediately notice
that consonants and vowels roughly alternate in their structure. When we
easily compress our text files on a computer, we remove such correlations
without losing information. Detailed analyses of DNA sequences have found
little correlation amongst the letters of its alphabet, and we have to
marvel at the fact that life has achieved the close to maximum entropy
structure of coding its information.

The third is the selection of the letters of the alphabet. This clearly
depends on the task to be accomplished and the choices available as symbols.
A practical criterion for fast error-free information processing is that
various symbols should be easily distinguishable from each-other. We use
the decimal system of numbers because we, at least in India, learnt to
count with our fingers. There is no way to prove this, but it is a better
explanation than anything else. We can offer a better justification for
why the computers we have designed use the binary system of numbers. 2 is
the smallest base available for a number system, and that leads to maximal
simplification of the elementary instruction set. (The difference is
obvious when we compare the mathematical tables we learnt in primary schools
to the corresponding operations in binary arithmetic.) 2 is also the maximum
number of items that can be distinguished with a single yes/no question.
(Detecting on/off in an electrical circuit is much easier than identifying
more values of voltages and currents.) Thus it is worth investigating what
life optimised in selecting the letters of its alphabet.

The computational task involved in DNA replication is ASSEMBLY. The desired
components already exist (they are floating around in a random ensemble);
they are just picked up one by one and arranged in the required order.
To pick up the desired component, one must be able to identify it uniquely.
This is a variant of the unsorted database search problem, unsorted because
prior to their selection the components are not arranged in any particular
order. (It is important to note that this replication is not a COPY process.
COPY means writing a specific symbol in a blank location, and unlike ASSEMBLY,
it is forbidden by the linearity of quantum mechanics.) The optimisation
criterion for this task is now clear---one must distinguish the maximum
number of items with a minimum number of identifying questions. I have
already pointed out that in a classical search, a single yes/no question
can distinguish two items. The interesting point is that a quantum search
can do better.

\section{\bf Unsorted Database Search}

Let the database contain $N$ distinct objects arranged in a random order.
A certain object has to be located in the database by asking a set of
questions. Each query is a yes/no question based on a property of the
desired object (e.g. is this the object that I want or not?). In the
search process, the same query is repeated using different input states
until the desired object is found. Let $Q$ be the number of queries
required to locate the desired object in the database.

Using classical probability analysis, it can be easily seen that
(a) $\langle Q \rangle = N$ when all objects are available with equal
probability for each query (i.e. each query has a success probability
of $1/N$), and
(b) $\langle Q \rangle = (N+1)/2$ when the objects which have been rejected
earlier in the search process are not picked up again for a query. Here
the angular brackets represent the average expectation values. Option (b)
is available only when the system possesses memory to recognise what has
already been tried before. In the random cellular environment, the rejected
object is thrown back into the database, and only option (a) is available
to a classical ASSEMBLY operation.

Lov Grover discovered a quantum database search algorithm that locates the
desired object using fewer queries \cite{grover}.
Quantum algorithms work with amplitudes, which evolve in time by unitary
transformations. At any stage, the observation probability of a state is
the absolute value square of the corresponding amplitude. The quantum
database is represented as an $N-$dimensional Hilbert space, with the $N$
distinct objects as its orthonormal basis vectors. The quantum query can
be applied not only to the basis vectors, but also to their all possible
superpositions (i.e. to any state in the Hilbert space). Let $|b\rangle$
be the desired state and $|s\rangle$ be the symmetric superposition state
\cite{notation}.
\begin{equation}
|b\rangle ~=~ (0 \ldots 0 1 0 \dots 0)^T ~~,~~
|s\rangle ~=~ (1/\sqrt{N}) (1 \ldots 1)^T ~~.
\end{equation}
Let $U_b = 1 - 2|b\rangle\langle b|$ and $U_s = 1 - 2|s\rangle\langle s|$
be the reflection operators corresponding to these states \cite{projection}.
The operator $U_b$ distinguishes between the desired state and the rest.
It flips the sign of the amplitude in the desired state, and is the query
or the quantum oracle. The operator $U_s$ treats all objects on an equal
footing. It implements the reflection about the average operation. Grover's
algorithm starts with the input state $|s\rangle$, and at each step applies
the combination $-U_sU_b$ to it. Each step just rotates the state vector by
a fixed angle (determined by $|\langle b|s\rangle| = 1/\sqrt{N}$) in the
plane formed by $|b\rangle$ and $|s\rangle$. $Q$ applications of $-U_sU_b$
rotate the state vector all the way to $|b\rangle$, at which stage the
desired state is located and the algorithm is terminated.
\begin{equation}
(-U_sU_b)^Q |s\rangle ~=~ |b\rangle ~~.
\end{equation}
This relation is readily solved, since the state vector rotates at a
constant rate, giving
\begin{equation}
(2Q+1) \sin^{-1} (1/\sqrt{N}) ~=~ \pi/2 ~~.
\end{equation}
Over the last few years, this algorithm has been studied in detail. I just
summarise some of the important features:
\begin{itemize}
\itemsep=0pt
\item   For a given $N$, the solution for $Q$ satisfying Eq.(3) may not be
an integer. This means that the algorithm will have to stop without the final
state being exactly $|b\rangle$ on the r.h.s. of Eq.(2). There will remain a
small admixture of other states in the output, implying an error in the search
process. The size of this admixture is determined by how close one can gets
to $\pi/2$ on the r.h.s. of Eq.(3). Apart from this, the algorithm is fully
deterministic.
\item   The algorithm is known to be optimal \cite{zalka},
going from $|s\rangle$ to $|b\rangle$ along a geodesic. No other algorithm,
classical or quantum, can locate the desired object in an unsorted database
with a fewer number of queries.
\item   The iterative steps of the algorithm can be viewed as the discretised
evolution of the state vector in the Hilbert space, governed by a Hamiltonian
containing two terms, $|b\rangle\langle b|$ and $|s\rangle\langle s|$. The
former represents a potential energy attracting the state towards $|b\rangle$,
while the latter represents a kinetic energy diffusing the state throughout
the Hilbert space. The alteration between $U_b$ and $U_s$ in the discretised
steps is reminiscent of Trotter's formula used in construction of the transfer
matrix from a discretised Feynman's path integral \cite{trotter}.
\item   Asymptotically, $Q = \pi\sqrt{N}/4$. The best that the classical
algorithms can do is to random walk through all the possibilities, and that
produces $Q = O(N)$ as mentioned above. With the use of superposition of all
possibilities at the start, the quantum algorithm performs a directed walk
to the final result and achieves the square-root speed-up.
\item   The result in Eq.(3) depends only on $|\langle b|s\rangle|$; the
phases of various components of $|s\rangle$ can be arbitrary, i.e. they can
have the symmetry of bosons, fermions or even anyons.
\end{itemize}

To come back to the genetic code, let us look at two of the solutions of
Eq.(3) for small $Q$. The only exact integral solution is $Q=1$, $N=4$.
Base-pairing during DNA replication can be looked upon as a yes/no query,
either the pairing takes place through molecular bond formations or it does
not, and its task is to distinguish between 4 possibilities. The other
interesting solution is $Q=3$, $N=20.2$. The well-known triplet code of
DNA has 3 consecutive nucleotide bases carrying 21 signals \cite{watson},
20 for the amino acids plus a STOP \cite{signals}.
3 base-pairings between t-RNA and m-RNA transfer this code to the amino acid
chain \cite{protsynth}.

These solutions are highly provocative. This is the first time they have
come out of an algorithm that performs the actual task accomplished by DNA.
It is fascinating that they are the optimal solutions \cite{errors}.
Indeed it is imperative to investigate whether DNA has the quantum hardware
necessary to implement the quantum search algorithm.

\section{\bf Molecular Biology and the Structure of DNA}

Over the last fifty years, molecular biologists have learnt a lot about the
structure and function of DNA by careful experiments (they have also been
rewarded with many Nobel prizes). Let us quickly go through some of the
facts they have unravelled \cite{watson,stryer,gribbin}.

\begin{figure}[tbh]
\epsfxsize=6cm
\centerline{\epsfbox{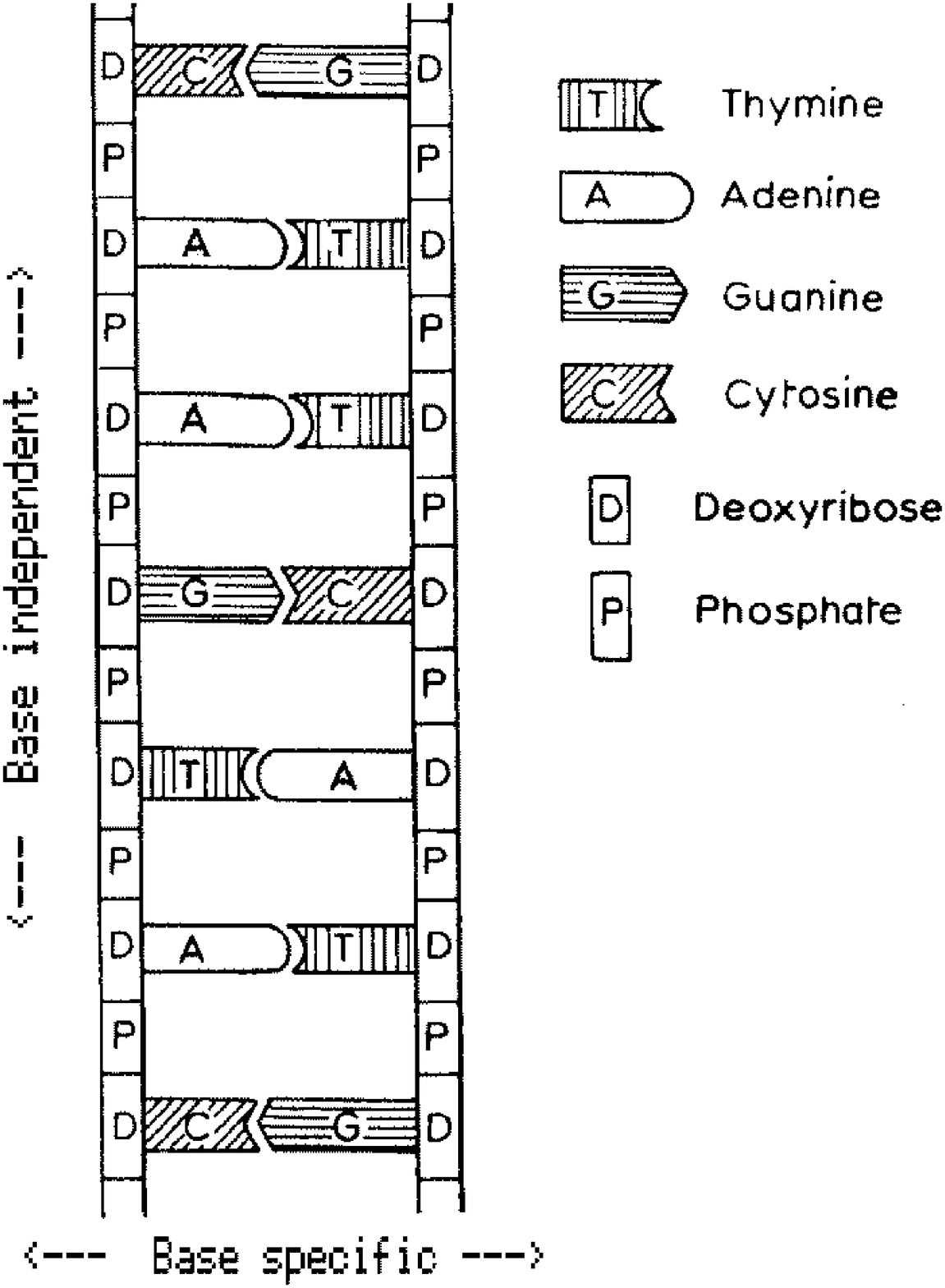}}
\caption{A schematic representation of the DNA double helix, depicting
base-specific and base-independent molecular bonds. The exact match between
bases joins two complementary strands of DNA.}
\end{figure}

\begin{itemize}
\itemsep=0pt
\item   DNA has the structure of a double helix. It can be schematically
represented as a ladder, as in Fig.1. The sides of the ladder have a periodic
structure with alternating sugar and phosphate groups. The nucleotide base
pairs form the rungs of the ladder, and the genetic information is encoded
in the order of these base pairs.
\item   DNA contains 4 nucleotide bases---A, T, C, G---which are closely
related to each-other in chemical structure \cite{binarybases}.
The bases are always paired as A-T and C-G along the rungs of the
ladder by Hydrogen bonds. This base-pairing makes the two DNA strands
complementary in character.
\item   The sugar and phosphate groups along the side of the ladder are
held together by covalent bonds. Their bonding is completely insensitive
to the bases attached to them, and takes place in the presence of DNA
polymerase enzymes.
\item   During replication, the helicase enzyme separates the two strands
of DNA by breaking the Hydrogen bonds, much like opening a zipper. The
unpaired bases along each of the separated strands find their partners
from the surrounding molecules, producing two copies of the original DNA.
\item   The sides of the ladder have asymmetric ends. The replication
process is directed, always proceeding from the $5'$ end to the $3'$ end
(these numbers label the position of the carbon atoms in the sugar rings)
of the strand being constructed. The DNA polymerase enzyme slides along
the intact strand, adding one base at a time to the growing strand.
During this process, base-pairing and sugar-phosphate bonding alternate.
\item   RNA molecules carry the nucleotide bases---A, U, C, G---with U
very similar in chemical structure to T. A-U pairing is as strong as A-T
pairing. Messenger RNA (m-RNA) has a single strand structure. In the first
step of protein synthesis, the RNA polymerase enzyme separates the paired
DNA strands and constructs an m-RNA strand on the DNA template by
base-pairing (the process is the same as in DNA replication). The m-RNA
strand grows as the RNA polymerase enzyme slides along the DNA from the
promoter to the terminator base sequence. Finally the RNA polymerase enzyme
detaches itself from the DNA, the fully constructed m-RNA strand floats away
to the ribosomes in the cytoplasm of the cell, and the separated DNA strands
pair up again.
\item   Transfer RNA (t-RNA) molecules have 3 RNA bases at one end and an
amino acid at the other, with a many-to-one mapping between the two. Inside
cellular structures called ribosomes, 3 t-RNA bases line up against the
matching bases of m-RNA, aligning the amino acids at the other end. The
aligned amino acids then split off from the t-RNA molecules and bind
themselves into a chain. The process again proceeds monotonically from the
$5'$ end to the $3'$ end of the m-RNA. After the amino acids split off,
the remnant t-RNA molecules are recycled. This completes the transfer of
the genetic code from DNA to proteins.
\item   Enzymes play a crucial role in many of the above steps. In addition
to facilitating various processes by their catalytic action, they store
energy needed for various processes, ensure that DNA keeps out U and RNA
keeps out T, and perform error correction by their $3'\rightarrow 5'$
exonuclease action
(i.e. reversing the assembly process to remove a mismatched base pair).
\item   Hereditary DNA is accurately assembled, with an error rate of
$10^{-7}$ per base pair, after the proof-reading exonuclease action.
Proteins are assembled less accurately, with an error rate of $10^{-4}$
per amino acid \cite{errors}.
\end{itemize}

Let us also recollect some useful facts about various chemical bonds.
\begin{itemize}
\itemsep=0pt
\item[$\circ$] Ionic bonds are strong, can form at any angle, and can be
explained in terms of electrostatic forces. Ions often separate in solutions.
\item[$\circ$] Covalent bonds are strong, form at specific angles, and
can be explained in terms of Coulomb forces between electrons and nuclei
and the exclusion principle.
\item[$\circ$] Van der Waals bonds are weak, not very angle dependent,
and explainable in terms of interactions between virtual electric dipoles.
They play an important role in transitions between solid, liquid and gas
phases, as well as in folding and linking of polymers.
\item[$\circ$] Hydrogen bonds are weak, highly angle dependent, and
explainable in terms of a proton ($H^+$) tunneling between two attractive
energy minima. The situation is a genuine illustration of a particle in a
double well potential, e.g. $O-H \cdots \ :N \Longleftrightarrow O^-:
\ \cdots H-N^+$. High sensitivity of the tunneling amplitude to the shape
of the energy barrier make Hydrogen bonds extremely sensitive to the
distances and angles involved. They are the most quantum of all bonds;
water is a well-known example.
\item[$\circ$] Delocalisation of electrons and protons over distances of
the order of a few angstroms greatly helps in molecular bond formation.
It is important to note that these distances are much bigger than the 
Compton wavelengths of the particles, yet delocalisation is common and
maintains quantum coherence. In case of electrons, the phenomena are
called resonance and hybridisation, e.g. the benzene ring. In case of
protons, the different configurations are called tautomers, e.g. amino
$\Leftrightarrow$ imino and keto $\Leftrightarrow$ enol fluctuations of
the nucleotide bases.
\end{itemize}

With all this information, the quantum search algorithmic requirements
from the DNA structure are clear. It is convenient to take the distinct
nucleotide bases as the quantum basis states in the Hilbert space. Then
(1) The quantum query transformation $U_b$ must be found in the
base-pairing with Hydrogen bonds.
(2) The symmetric transformation $U_s$ must be found in the
base-independent processes occurring along the sides of the ladder.
(3) An environment with good quantum coherence must exist.
Thermal noise is inevitable at $T \approx 300^\circ K$ inside the cells,
so the transformations must be stable against such fluctuations.
Figuratively, the best that can be achieved is
\begin{equation}
{\rm Actual~evolution} ~=~ {\lim \atop {\rm decoherence}\rightarrow 0}
                           [ {\rm Quantum~evolution} ] ~~.
\end{equation}
Thus we need quantum features that smoothly cross over to the classical
regime, i.e. features that are reasonably stable against small decoherent
fluctuations. Examples are: (a) geometric and topological phases, and
(b) projection/measurement operators.

\section{\bf Base-pairing as the Quantum Oracle}

During DNA replication, the intact strand of DNA acts as a template on
which the growing strand is assembled. At each step, the base on the
intact strand decides which one of the four possible bases can pair with
it. This is exactly the yes/no query (also called the oracle) used in
the database search algorithm. To connect this oracle to the quantum
transformation $U_b$, we have to look at how molecular bond formations
transform a quantum state.

The generic quantum evolution operator is $\exp(-iHt)$, with $H$ being
the total Hamiltonian of the system. Global conservation of energy means
that the overall phase $\exp(-iEt)$ will completely factor out of the
evolution and will not affect the final probabilities. What matters is
only the relative phase between pairing and non-pairing bases. During the
pairing process, the bases come together in an initial scattering state,
discover that there is a lower energy binding state available, and decay
to that state releasing the extra energy as a quantum \cite{energy}.
The interaction Hamiltonian for the bond formation process can be
represented as
\begin{equation}
H_{int} \propto (a^\dagger b ~+~ b^\dagger a) ~~,
\end{equation}
where $a,a^\dagger$ are the transition operators between the excited
and ground states of the reactants, and $b,b^\dagger$ are the transition
operators between zero and one quantum states of energy released. Both
the terms in Eq.(5) are necessary for the Hamiltonian to be Hermitian.
The phase change $\varphi$ during the bond formation satisfies
\begin{equation}
\exp(-iH_{int}t_b) |e\rangle |0\rangle ~=~ \varphi |g\rangle |1\rangle ~~.
\end{equation}
With only two states involved, Eq.(6) is easily solved by diagonalising
$H_{int}$. In the two dimensional space, let 
\begin{equation}
H_{int} ~=~ \Delta E_H \left( \matrix{0&1 \cr 1&0 \cr} \right) ~~,~~
                      {\rm eigenvectors:}~{1 \over \sqrt{2}}
                       \left( \matrix{1 \cr \pm1 \cr} \right) ~~,
\end{equation}
and eigenvalues $\pm \Delta E_H$. Eq.(6) then reduces to
\begin{equation}
\exp(-i\Delta E_H t_b) ~=~ \varphi ~=~ -\exp(i\Delta E_H t_b) ~~,
\end{equation}
with the solution $\varphi = \sqrt{-1}$.

This is the geometric phase well-known in quantum optics. A complete Rabi
cycle in a two-level system gives a phase change of $-1$, and the transition
process corresponds to half the cycle. The phase $\varphi$ does not depend
on specific values of $\Delta E_H$ or $t_b$, but only on the fact that the
transition takes place. Quantum mechanics does not specify how $\varphi$
will be divided between the bound state and the released energy quantum.
In quantum optics, this break-up is determined by the phase of the laser.
Here I assume that the process of decoherence is such that the energy
quantum does not carry away any phase information.

At this stage, we discover a pleasant surprise that the base-pairing takes
place not with a single Hydrogen bond but with multiple Hydrogen bonds (two
for A-T and A-U, and three for C-G), as shown in Fig.2 \cite{twobonds}.
Multiple Hydrogen bonds are necessary for the mechanical stability of the
helix. But they are also of different length, making it likely that they form
asynchronously. Assuming a two-step deexcitation process during base-pairing,
the geometric phase change becomes $\varphi^2=-1$, just what is needed to
implement $U_b$.

The energy of a single Hydrogen bond is $\Delta E_H \approx 7 kT$, giving 
$\exp(-\Delta E_H/kT) \sim 10^{-3}$. This roughly explains the observed
error rate in DNA replication. The tunneling amplitude for bond formation
is related to both $\Delta E_H$ and $t_b$, which determines the time scale
of the base-pairing,

\begin{figure}[tbh]
\epsfxsize=8cm
\centerline{\epsfbox{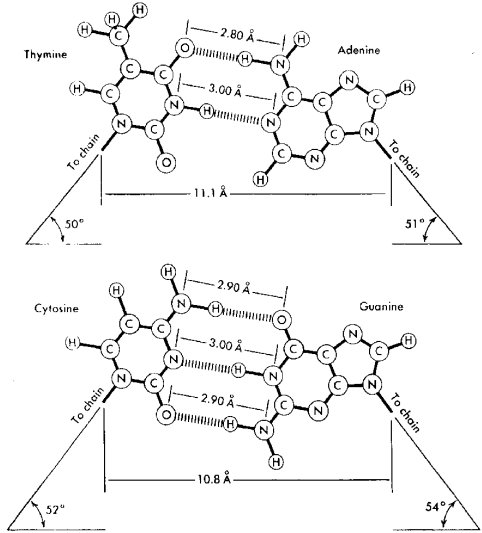}}
\caption{The base-pairing in DNA with Hydrogen bonds. The figure is taken
from Ref.[12].}
\end{figure}

\begin{equation}
\Delta E_H t_b ~\approx~ \hbar ~~\Longrightarrow~~
t_b ~\approx~ 4 \times 10^{-15}~{\rm sec}.
\end{equation}

\section{\bf A Quantum Search Scenario}

The next step is to look for the transformation $U_s$ in the processes
occurring along the sides of the DNA-ladder. During these processes, quantum
evolution produces various phases as molecular bonds get formed and broken.
But these bonds treat all the nucleotide bases in the same manner, so the
phases completely factor out and have no effect on the final probabilities.
Thus I leave the phases out.

Suppose that $|s\rangle$ is the equilibrium state of the physical system,
favoured by the processes that occur along the sides of the DNA-ladder.
This means that any other initial state will gradually relax towards
$|s\rangle$, with the damping provided by the environment. Let $t_r$
be the time scale for this relaxation process. Now $|s\rangle$ is a
superposition state of nucleotide bases, and it can be created only if
the cellular environment provides transition matrix elements between its
various components. (In free space, transition matrix elements between
nucleotide bases of different chemical composition vanish.) The magnitude
of these transition matrix elements decides how quickly $|s\rangle$ cycles
through its various components. Let $t_{osc}$ be the time scale for these
oscillations. Now let us look at the DNA replication process, when the above
defined time scales satisfy the hierarchy
\begin{equation}
t_b ~~<<~~ t_{osc} ~~<< t_r ~~.
\end{equation}

\vspace{-0.4truecm}
\begin{enumerate}
\item[(1)] In the initial stage, the randomly floating around nucleotide
bases come into contact with the growing DNA strand, and relax to the state
$|s\rangle = (1/\sqrt{N}) \sum_i |i\rangle$.
\item[(2)] When the nucleotide base finds its proper orientation, Hydrogen
bond formation suddenly takes place, changing the state to $U_b|s\rangle$.
This state is entangled between the nucleotide bases and the energy quanta,
$U_b|s\rangle = (1/\sqrt{N})
[ \sum_{i \ne b} |i\rangle|0\rangle - |b\rangle|2\rangle ]$.
\item[(3)] After this sudden change, the relaxation process again tries to
bring the system back to the state $|s\rangle$. With the time scales
obeying Eq.(10), this relaxation occurs as damped oscillations, much like
what happens when one gives a sudden jerk to a damped pendulum.
\item[(4)] The opposite end of the damped oscillation is
$(2|s\rangle\langle s|-1)U_b|s\rangle = -U_sU_b|s\rangle$.
When the system evolves to this opposite end, it discovers that it is no
longer entangled between the nucleotide bases and the energy quanta,
$-U_sU_b|s\rangle|0\rangle = |b\rangle|2\rangle$ for $N=4$. At this point,
the energy quanta are free to wander off with minimal disturbance to the
quantum coherence. The departure of the energy quanta confirms the
base-pairing, providing a projective measurement of the system.
(I take the measurement time scale to be much smaller than $t_{osc}$.)
\item[(5)] The energy quanta that have wandered off are unlikely to return,
making the process irreversible. The replication then continues to add the
next base onto the growing strand.
\end{enumerate}

\begin{figure}[tbh]
\epsfxsize=8cm
\centerline{\epsfbox{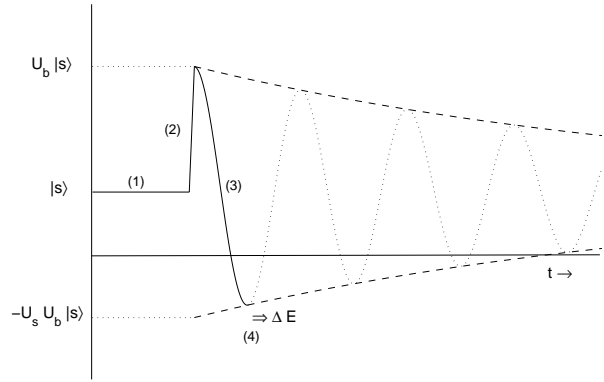}}
\caption{A time evolution picture of the DNA replication algorithm. Numerical
labels refer to the steps described in the text.}
\end{figure}

These steps are schematically shown in Fig.3. They provide a highly tuned
yet a robust algorithm. There are no fine-tuned parameters; the hierarchy
of Eq.(10) has enough room to take care of substantial variation in
individual time scales. Life couldn't be simpler!

To illustrate the possibility that processes with time scales obeying the
hierarchy of Eq.(10) do physically occur, let us look at a quantum example.
Consider the processes involved in the functioning of the $NH_3$ maser.
$NH_3$ is a molecule with two equivalent configurations, corresponding to
the Nitrogen atom being above or below the triangle of Hydrogen atoms.
These two configurations can be distinguished by the direction of their
electric dipole moments. Quantum tunneling of the Nitrogen atom through
the triangle of Hydrogen atoms mixes these two configurations, and the
ground state is the symmetric superposition of the two. In $NH_3$ gas,
any initial state decays towards this equilibrium ground state (molecular
collisions help in this relaxation). If a electric field is applied to the
gas for a short duration, it favours one of the two configurations and
kicks the molecules out of their equilibrium state. After the removal of
the electric field, the kicked molecules oscillate from one configuration
to the other, till the oscillations are damped out by the decay process.
Shining the molecules with a radio-frequency pulse resonant with $t_{osc}$
removes the extra energy quickly by stimulated emission and produces a
coherent maser.

The relaxation time scale $t_r$ depends on temperature as well as on various
molecular concentrations, and governs the overall replication rate. Under
normal circumstances, DNA replication is observed to occur at the rate of
1000 base-pairings/sec, constraining $t_r$ to be smaller than $O(10^{-3})$
sec. Without any knowledge of the transition matrix elements, I do not have
any estimate of $t_{osc}$.

Strictly speaking, I should not talk about quantum states in processes
that involve damping; the proper mathematical formulation must be in the
language of density matrices. But when the damping is small, as is the
case here, it is easier to talk about states. The steps above can be easily
transcribed in the language of density matrices, without changing their
outcomes.

Now we can proceed to the remaining pieces needed to complete the scenario:
a mechanism that favours the state $|s\rangle$ as the equilibrium state,
and an environment that permits an almost coherent quantum evolution.
For that I have to appeal to the ingredients ignored so far---the enzymes.

\section{\bf The Role of the Enzymes}

Enzymes play a very important role in many biochemical reactions, and some
of the things they do were mentioned in section III. The rates of various
biochemical reactions, when estimated with the standard thermodynamical
analysis (probability distributions, diffusion processes, kinetic theory,
etc.), fall too short of the observed rates, often by orders of magnitudes.
One is left with no choice but to admit that these reactions are catalysed.

Enzymes are the objects which catalyse biochemical reactions. They are large
complicated molecules, much larger than the reactants they help, made of
several peptide chains. Their shapes play an important part in catalysis,
and often they completely surround the reaction region. They do not bind
to either the reactants or the products, just help them along the way. The
standard explanation is that enzymes lower reaction barriers. Just how this
lowering of reaction barriers occurs is not clearly understood, and is an
active area of research (for example, enzymes can form weak bonds with the
transition state or suck out solvent molecules from in between the reactants).
Ultimately it must be explained in terms of some underlying physical laws.

I put forward two specific hypotheses about what enzymes accomplish in the
processing of genetic information.
\begin{itemize}
\itemsep=0pt
\item   {\it Enzymes provide a shielded environment where quantum coherence
of the reactants is maintained.} This is a rather passive task, consistent
with the properties of enzymes mentioned above, and it is also plausible.
For instance, diamagnetic electrons do an extraordinarily good job of
shielding the nuclear spins from the environment---the coherence time
observed in NMR is $O(10)$ sec, much longer than the thermal environment
relaxation time ($\hbar/kT \sim O(10^{-14})$ sec) and the molecular collision
time ($O(10^{-11})$ sec), and still neighbouring nuclear spins couple through
the electron cloud. A few orders of magnitude increase in coherence time is
sufficient in many reactions for faster quantum algorithms to take over from
their classical counterparts, and provide the catalytic speed-up.
\item   {\it Enzymes are able to create superposed states of chemically
distinct molecules.} This is an active task. Various nucleotide bases differ
from each other only in terms of small chemical groups, containing less than
10 atoms, at their Hydrogen bonding end. To convert one base into another,
enzymes have to be repositories of these chemical groups which differentiate
between various nucleotide bases. Enzymes are known to do cut-and-paste jobs
with such chemical groups (e.g. one of the simplest substitution processes
is methylation, replacing $-H$ by $-CH_3$, which converts U to T). Given
such transition matrix elements, quantum dynamics automatically produces a
superposition state as the lowest energy equilibrium state. (Note that the
cut-and-paste job in a classical environment would produce a mixture, but
in a quantum environment it produces superposition.) It is mandatory that
the enzymes do the cut-and-paste job only on the growing strand and not on
the intact strand. Perhaps this is ensured by other molecular bonds.
\end{itemize}

These hypotheses are powerful enough to explain many observed properties of
enzymes from a new perspective. For example,
\begin{itemize}
\itemsep=0pt
\item[$\circ$] It is obvious why DNA replication always takes place in the
presence of enzymes. If base-pairing were to occur by chance collisions,
it would occur anywhere along the exposed unpaired strand.
\item[$\circ$] Since enzymes control the transition matrix elements, they
can keep U out of DNA and T out of RNA.
\item[$\circ$] Quantum processes can tunnel through the reaction barriers,
instead of climbing over them.
\item[$\circ$] As long as quantum coherence is maintained, the replication
process is reversible. This can easily explain the error-correcting
exonuclease action of the polymerase enzymes.
\end{itemize}

More importantly, these hypotheses are experimentally testable, provided
one can observe the replication process at its intermediate stages. That
is within the grasp of modern techniques such as X-ray diffraction analysis,
electron microscopy, NMR spectroscopy, radioactive tagging of atoms in
chemical reactions, and femtosecond photography.

\section{\bf Summary and Future}

I have proposed a quantum algorithmic mechanism for DNA replication and
protein synthesis. This genetic information processing takes place at the
molecular level, where quantum physics is indeed the dominant dynamics
(classical physics effects appear as decoherence and are subdominant).
It is reasonable to expect that if there was something to be gained from
quantum computation, life would have taken advantage of that at this
physical scale \cite{neurons}.
For DNA replication, the quantum search algorithm provides a factor of two
speed-up over classical search, but it is still an advantage. Quantum
algorithms can provide a bigger advantage for more complicated processes
involving many steps.

Implementation of quantum database search does not require a general purpose
quantum computer. A system that can implement the quantum oracle and quantum
state reflections is sufficient. I have described the physical analogues of
these operations in genetic information processing, and it is not unusual
for living organisms to find the correct ingredients without bothering about
generalities. In fact various pieces of the scenario have fitted together
so nicely (database search paradigm, optimal $(Q,N)$ values---$(1,4)$ and
$(3,20.2)$, quantum transformations implementing $U_b$ and $U_s$),
that with the courage of conviction, I have made bold hypotheses to fill the
remaining gaps. The role that enzymes play, as described in section VI, is
both plausible and experimentally verifiable. The other assumptions I have
made (i.e. two-step cascade deexcitation in base-pairing, energy quanta not
carrying away any phase information, energy quanta departing only when they
cause minimal decoherence to the system) concern how decoherence modifies
pure quantum evolution, and are also experimentally testable.
Such experimental tests would decide the future of this proposal.

It is clear that if experiments verify the quantum scenario for genetic
information processing presented here, there will be a significant overhaul
of conventional molecular biology. There is nothing inappropriate in
that---the subject of molecular biology was born out of quantum physics,
and quantum physics has not yet had its last word on it.

I also want to acknowledge that molecular biology is a far more complicated
subject than the simplest features I have explored in this work. If quantum
physics provides better explanations for other more complex phenomena of
life, it would be a wonderful development.

\section*{\bf Appendix: Some Questions and Answers}

I am grateful to the audience for their many comments and questions.
I summarise below some of the important concepts they brought out.

\noindent Q: Are there any examples of genetic codes which use other values
of $Q$, say $Q=2$?

\noindent A: I do not know of any such examples. May be some day we shall
become clever enough to synthesise such instances in the laboratory.

\noindent Q: Parallel processing can also speed-up algorithms. Why is that
not used in the case of genetic code?

\noindent A: In case of DNA replication, enzymes separate only a limited
region of paired strands as they slide along replicating the information.
Complete separation of the strands would break too many Hydrogen bonds and
cost too much in energy. Thus the replication remains a local process. In
case of protein synthesis, random pairing of t-RNA and m-RNA at several
different locations would lead to a lot of mismatches and errors, since the
triplet code needs precise starting and ending points. Parallel processing
does take place though: several ribosomes work simultaneously on a single
m-RNA strand, each one traversing the full length of m-RNA from one end to
the other and constructing identical amino acid chains.

\noindent Q: Three nucleotide bases can form 64 distinct codes. Why do you
still say that 20 is the optimal number for the triplet code?

\noindent A: When the DNA assembly takes place, with each base as a separate
unit, the number of possibilities explored is indeed $4^Q$. But this is not
what happens in the matching between t-RNA and m-RNA. The three bases come
as a single group, without any possibility of their rearrangement. The whole
group has to be accepted or rejected as a single entity. In such a situation,
the number of objects that can be distinguished is smaller, as given by Eq.(3).

\noindent Q: Do you have any understanding of degeneracy of the triplet code?

\noindent A: With only 21 signals embedded amongst 64 possibilities, the
amino acid code is indeed degenerate. Moreover, all the amino acids are not
present in living organisms with equal frequency, quite unlike the nucleotide
bases. Ribosomes are also much bigger and more complicated molecules than the
polymerase enzymes, and so capable of carrying out more complex tasks. All
this makes protein synthesis a more difficult process to study than DNA
replication. I have not analysed it in detail, and I am unable to provide any
understanding of the degeneracy of the triplet code.

\noindent Q: The energy quanta do not have to be released exactly at the
opposite end of the oscillation. Oscillations spend more time near their
extrema anyway, and so even random emission will give high probability for
correct base-pairing.

\noindent A: I agree. But a smaller success probability means that more
trials will be necessary to achieve the correct base-pairing. That would
diminish the advantage of the quantum algorithm.

\noindent Q: Enzymes are also proteins which have to be synthesised by the
DNA. How are they synthesised in the first place?

\noindent A: Classical algorithms can do everything that quantum algorithms
do, albeit slower. In the absence of enzymes, various steps will take place
only by random chance, and so the start-up will be slow. But once processes
get going, as in a living cell, enzymes will be manufactured along the way
and there will be no turning back to slower algorithms.

\noindent Q: If polymerase enzymes have to keep on supplying various chemical
groups to nucleotide bases, they would run out of their stock at some stage.
What happens then?

\noindent A: In my proposal, the polymerase enzymes substitute one chemical
group for another. With the DNA base sequence being random, the chemical
groups are recycled with high probability. So with a reasonable initial stock
an enzyme can perform its task for a long time. Of course if an enzyme runs
out of its stock, it has no choice but to quit and replenish its stock.

\noindent Q: How essential is the environment of a living cell in DNA
replication? Are there any other molecules besides enzymes involved?

\noindent A: DNA replication can take place without cellular environment.
With proper enzymes in the solution, the polymerase chain reaction can
rapidly multiply DNA. This is used in DNA fingerprinting from dead cells.

\noindent Q: Many inorganic reactions can be speeded up by appropriate
catalysts. Are they quantum reactions too?

\noindent A: In my view, catalysts convert random walk processes into
directed walks, providing a $\sqrt{N}$ speed-up in the number of steps.
Quantum superposition is one way to achieve this, but it does not have to
be the only way. There may be even classical mechanisms which can do the
same job.

\noindent Q: Living organisms are known to emit radiation, which is coherent
and not thermal (black-body). Is there any connection between these biophotons
and your proposal?

\noindent A: This is the first time I have heard about biophotons. If they are
related to some quantum processes going on inside living cells, that would be
great.

\noindent Q: Are there any applications of your proposal?

\noindent A: Understanding the basic processes of life will always lead
to new applications. For example, molecular biologists have been working
on accelerating synthesis of desired proteins and inhibiting growth of
cancer and harmful viruses. I am not inclined to speculate more on this
issue right now.


\begin{thebibliography}{99}

\bibitem{lifeshapes}{I mention here the explanation of physical shapes of
living organisms \cite{gardner}.
At microscopic level, the physical forces (e.g. diffusion, surface tension)
are essentially isotropic. Consequently single free cells are found to be
more or less spherical in shape (e.g. many bacteria, amoeba). At larger scales,
the isotropy is explicitly broken by gravity. This leads to large stationary
living organisms being axially symmetric about the vertical direction (e.g.
plants, hydra). The symmetry is also broken spontaneously by the need of the
organisms to move/grow in order to find food. For small enough organisms where
the effects of gravity is not important, there arises an axial symmetry about
the direction of movement (e.g. worms, some bacteria). For most animals, both
gravity and movement are important, with gravity restricting the direction
of movement to be horizontal. The surviving symmetry is then bilateral, i.e.
mirror reflection in the plane formed by directions of gravity and movement.
What is striking in this example is the simplicity of the logic, i.e. how far
one can go with how little input once the ingredients are right. It is also
obvious that there is a limit to how far simple logic can be pushed---to
explain the structure in more depth, one would require more detailed physical
data.}

\bibitem{gardner}{M. Gardner, {\it The New Ambidextrous Universe:
	 Symmetry and Asymmetry, from Mirror Reflections to Superstrings},
         Third Edition, W.H. Freeman (1991).}

\bibitem{chirality}{For the above explanation of shapes of living organisms,
one can estimate the scale at which effects of gravity and motion would
become important. The orders of magnitude estimates do come out right.
On the other hand, many attempts have been made to link observed chirality
of important molecules of life (sugars and amino acids) to the chirality of
the weak interactions. But weak interactions are just too weak to produce
any influence at the molecular scale, and no attempt has come anywhere close
to convincing.}

\bibitem{perpetuation}{The same physical laws used by living organisms for
self-perpetuation, also imply that the atoms making up the organisms will
last forever. In that case, why the atoms should first organise themselves
in complicated structures, and then try to perpetuate them, is something
that I cannot answer.}

\bibitem{protsynth}{Protein synthesis involves several steps. First the
DNA synthesises messenger RNA molecules, much in the same manner that it
replicates itself. These m-RNA molecules then travel from the nucleus of the
cell to the ribosomes in its cytoplasm. There, with the help of transfer RNA
molecules, they construct proteins from amino acids. The matching between
m-RNA and t-RNA molecules follow the same rules as in DNA replication.}

\bibitem{schrodinger}{E. Schr\"odinger, {\it What is Life?}, Cambridge
University Press (First published 1944).}

\bibitem{grover}{L. Grover, {\it A Fast Quantum Mechanical Algorithm for
Database Search}, Proceedings of the 28th Annual ACM Symposium on Theory
of Computing, Philadelphia (1996), p.212, {\tt quant-ph/9605043}.}

\bibitem{notation}{I am using the standard quantum mechanical notation
introduced by Dirac. The suffix $b$ is used, since that will correspond
to the state which forms Hydrogen bonds in nucleotide base-pairing.}

\bibitem{projection}{A projection operator $P = |x\rangle\langle x|$
satisfies $P^2 = P$. Then $(1-P)$ is also a projection operator,
$(1-P)^2 = 1-P$, and $(1-2P)$ is a reflection operator, $(1-2P)^2 = 1$.}

\bibitem{zalka}{C. Zalka, {\it Grover's quantum searching algorithm is
optimal}, Phys. Rev. A60 (1999) 2746, {\tt quant-ph/9711070}.}

\bibitem{trotter}{L. Grover, these proceedings.}

\bibitem{watson}{J.D. Watson, N.H. Hopkins, J.W. Roberts, J.A. Steitz and
A.M. Weiner, {\it Molecular Biology of the Gene}, Fourth Edition,
Benjamin/Cummings (1987).}

\bibitem{signals}{The START signal for synthesis of an amino acid chain
is somewhat complicated; it is represented by the code for the amino acid
Methionine preceded by an arrangement of several bases. But the STOP signal
for terminating the chain is encoded in the DNA base sequence in the same
manner as the signals for specific amino acids.}

\bibitem{errors}{DNA replication should be error-free to faithfully pass
on the hereditary information to the next generation. Protein synthesis
can tolerate more errors, since proteins are disposable items, made in
large numbers and broken up into parts once their use is over. With $Q=3$
and $N=21$, the quantum search algorithm has an error rate of about 1 part
in 1000.}

\bibitem{stryer}{L. Stryer, {\it Biochemistry}, Fourth Edition,
W.H. Freeman (1995).}

\bibitem{gribbin}{An informative book at popular level is: J. Gribbin,
{\it In Search of the Double Helix: Quantum Physics and Life}, Penguin
Books (1995).}

\bibitem{binarybases}{It is easy to give the bases binary labels based on
their chemical structure. The first bit can be a pyrimidine/purine label
(i.e. a single or a double ring), and the second bit an amino/keto label
(i.e. $-NH_2$ or $=O$ group). Then pairing occurs amongst bases differing
in both the bits.}

\bibitem{energy}{In the cellular environment, it is far more likely that
this binding energy is released as an inelastic collision or a phonon,
than the radiation of a photon.}

\bibitem{twobonds}{A different nucleotide base, I, appears sometimes in
the wobble position of t-RNA. In the base-pairing between t-RNA and m-RNA
at this position, the additional combinations, I-A, I-C, I-U and U-G, are
held together by two Hydrogen bonds.}

\bibitem{neurons}{We can also look at a different example. Living organisms
also carry out information processing using their nervous systems. This
processing takes place at the cellular level, where classical physics
dominates. Expectedly, neural communications use binary code---neurons
communicate with each-other by firing electrical pulses, and a neuron fires
or does not fire depending on whether its input potential is above or below
certain threshold.}

\end{thebibliography}
\end{document}